\documentclass[%
aps,
pra,
amsmath,amssymb,
reprint,
showpacs,
superscriptaddress,
]{revtex4-1}
\usepackage{graphicx} 
\pdfoutput=1

\usepackage{txfonts}
\def\bvec#1{{\mathchoice{\mbox{\boldmath$#1$}}{\mbox{\boldmath$#1$}}%
  {\mbox{\scriptsize\boldmath$#1$}}{\mbox{\scriptsize\boldmath$#1$}}}}

\makeatletter
\AtBeginDocument{\@ifpackageloaded{natbib}{\ifNAT@numbers\if@filesw\immediate\write\@auxout{\string\global\string\NAT@numberstrue}\fi\fi}{}}
\makeatother
\begin{document}

\title{Observation of a nonradiative flat band for spoof surface plasmons in a metallic Lieb lattice}
\author{Sho Kajiwara}
\email{kajiwara@giga.kuee.kyoto-u.ac.jp}
\affiliation{Department of Electronic Science and Engineering, Kyoto University, Kyoto 615-8510, Japan}
\author{Yoshiro Urade}
\affiliation{Department of Electronic Science and Engineering, Kyoto University, Kyoto 615-8510, Japan}
\author{Yosuke Nakata}
\affiliation{Center for Energy and Environmental Science, Shinshu University, 4-17-1 Wakasato, Nagano 380-8553, Japan}
\author{Toshihiro Nakanishi}
\affiliation{Department of Electronic Science and Engineering, Kyoto University, Kyoto 615-8510, Japan}
\author{Masao Kitano}
\email{kitano@kuee.kyoto-u.ac.jp}
\affiliation{Department of Electronic Science and Engineering, Kyoto University, Kyoto 615-8510, Japan}
\date{\today}
\begin{abstract}
We demonstrate a nonradiative flat band for spoof surface plasmon polaritons bounded on a structured surface with Lieb lattice symmetry in the terahertz regime. 
First, we theoretically derive the dispersion relation of spoof plasmons in a metallic Lieb lattice based on the electrical circuit model. 
We obtain three bands, one of which is independent of wave vector. 
To confirm the theoretical result, we numerically and experimentally observe the flat band in  transmission and attenuated total reflection configurations. 
We reveal that the quality factor of the nonradiative flat band mode decoupled from the propagating wave is higher than that of the radiative flat band mode. 
This indicates that the nonradiative flat band mode is three-dimensionally confined in the lattice. 
\end{abstract}
\maketitle
%
%
\section{Introduction}\label{sec1}
\vspace{-2mm}
In the terahertz (THz) region and lower frequency regions, surface
plasmon polaritons confined at a metal-dielectric interface cannot exist
because of the high conductivity of metals \cite{maier2007plasmonics}.
However, if the surface of the metal is artificially structured,
electromagnetic modes similar to the surface plasmons can be realized \cite{maier2006terahertz, williams2008highly, pendry2004mimicking}.
These surface-plasmon-like modes are called spoof surface plasmon polaritons. 
The dispersion relation of spoof surface plasmon polaritons can be
controlled  by appropriately designing the structure of the metal
surface \cite{maier2006terahertz, martin2008spoof, gan2008ultrawide, yu2010designer}.
Especially, if the group velocity of the spoof surface plasmon polaritons is slowed down, the plasmon-matter interaction is enhanced as is the case in light-matter interaction. 
To realize the slow group velocity of the spoof surface plasmon
polaritons, we focus on the fact that the electron dispersion relation
in  crystals with specific lattice structures, such as Lieb-type
\cite{PhysRevLett.62.1201}, Tasaki-type \cite{tasaki1992ferromagnetism},
and Mielke-type \cite{mielke1991ferromagnetism} is independent of wave number and the group velocity becomes zero in any direction owing to destructive interference. 
The wave-number-independent band is called a flat band. 
Recently, flat bands for electromagnetic waves have been realized in
photonic crystals \cite{takeda2004flat, vicencio2015observation, PhysRevLett.114.245504}.
They are also expected to exist in metallic waveguide networks
\cite{endo2010tight, feigenbaum2010resonant}.

In  previous work \cite{nakata2012circuit}, utilizing electrical circuit models, we showed that the dispersion relations of metal structures with various lattice symmetries have flat bands. 
We confirmed the formation of the electromagnetic flat bands for a
metallic kagom$\acute{\rm e}$ lattice by illumination of propagating
waves theoretically and experimentally \cite{nakata2012observation}.
In this case, however, the flat-band mode is only two-dimensionally localized in the in-plane directions and the energy escapes into  free space. 
This is because the flat band is located inside the light cone or in a radiative region. 
Thus the mode, which is coupled with free-space modes, is considered as spoof plasmon polaritons, not spoof {\it surface} plasmon polaritons. 
For complete three-dimensional confinement of spoof surface plasmon polaritons, we have to implement a flat band outside the light cone or in a nonradiative region decoupled from free-space modes. 
In this paper, we demonstrate that the flat band for spoof surface plasmon polaritons can be realized in a metallic Lieb lattice. 
In the metallic Lieb lattice, the nonradiative flat-band mode appears in lower frequencies than in the metallic kagom$\acute{\rm e}$ lattice for the same design parameters, and the nonradiative flat-band region is broadened compared with that of the kagom$\acute{\rm e}$ lattice. 
The nonradiative flat-band mode for spoof surface plasmon polaritons is excited by illumination of evanescent waves. 

This paper is organized as follows: 
In Sec.~\ref{sec2}\@, we theoretically derive the dispersion relation of spoof plasmons in Lieb-type bar-disk resonators based on an electrical circuit model. 
In Sec.~\ref{sec3}\@, we numerically calculate the electromagnetic response of Lieb-type bar-disk resonators using a finite-element method. 
In Sec.~\ref{sec4}\@, we describe our experimental setup for the observation of the flat band and  show the experimental results and compare them with the simulation results. 
In Sec.~\ref{sec5}\@, we discuss the results in detail. 
Finally, in Sec.~\ref{sec6}\@, we conclude our study and propose some applications of the nonradiative flat-band mode.

\section{theoretical model}\label{sec2}
\begin{figure}[t]
\includegraphics[scale=0.19]{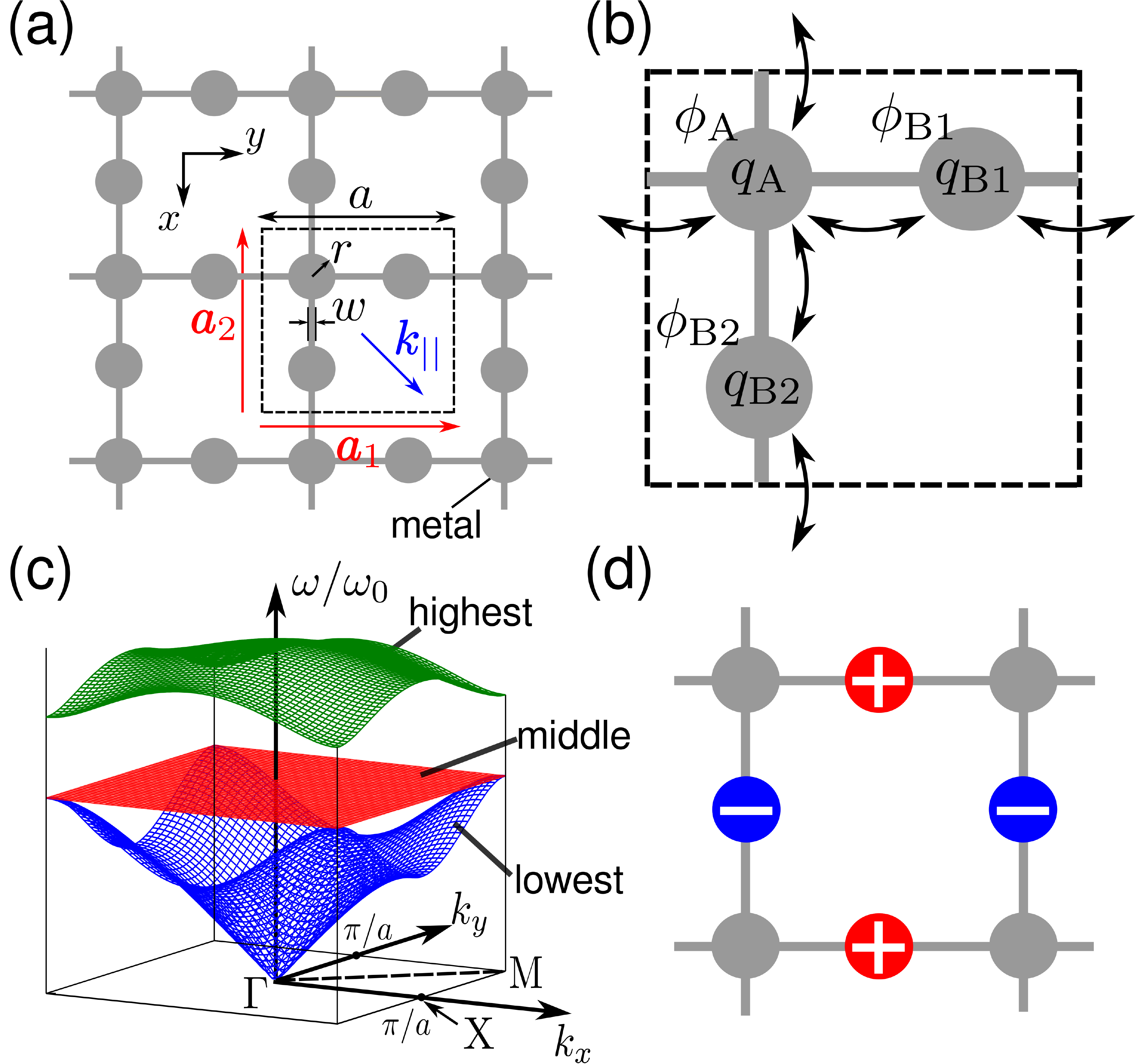}
\caption{\label{fig1} (Color online) (a) Design of the LBDRs: Length of a unit cell, $a$, radius of a disk, $r$, and width of a bar, $w$. The wave vector $\bvec{k}_{||}$ on a plane of the LBDRs is depicted. The vectors $\bvec{a}_{1}$ and $\bvec{a}_{2}$ are unit lattice vectors. (b) A unit cell of the LBDRs: $q_{k}$ ($k={\rm A,\,B1,\,B2}$) is the electric charge on the disk, and $\phi_{k}$ is the electric potential at the disk. The arrows indicate the nearest-neighbor capacitive couplings. (c) Dispersion relation of the LBDRs for $\eta=0$. The middle band is flat at $\omega/\omega_{0}=\sqrt{2}$. (d) Charge distribution of a localized mode forming the flat band. }
\vspace{-5mm}
\end{figure}

\vspace{-2mm}
Using a model similar to the one in our previous study \cite{nakata2012observation}, we derive the dispersion relation for spoof plasmons in the Lieb-type bar-disk resonators (LBDRs), as shown in Fig.~\ref{fig1}(a). 
The lattice formed by the disks connected to four disks is named the {\it main lattice}, and the lattice formed by the disks connected to two disks is named the {\it sublattice}. 
We assume that the structure is made of an infinitely thin, lossless metal and the electric charge is stored only on each disk and oscillates alternately. 
The electric potential $\phi_{i}$ of the $i$th disk can be expressed with the electric charge $q_{j}$ on the $j$th disk as follows: 
\begin{align}\label{1}
\phi_{i}=Uq_{i}+U'\sum_{j}A_{ij}q_{j}. 
\end{align}
We only consider the self- and  nearest-neighbor capacitive couplings, as shown in Fig.~\ref{fig1}(b). 
The constants $U$ and $U'$ are the potential coefficients, and $A_{ij}$ is an adjacency matrix of the lattice; i.e.,  $A_{ij}=1$ if the $i$th and $j$th disks are connected directly by a bar and otherwise $A_{ij}=0$. 
Using the wave vector $\bvec{k}_{||}$ on the plane of the LBDRs and the unit-lattice vectors $\bvec{a}_{1}$ and $\bvec{a}_{2}$ shown in Fig.~\ref{fig1}(a), we can obtain the relation between the electric potentials and the electric charges of the three disks (A, B1, and B2) in the unit cell as follows: 
\begin{align}\label{2}
\small
\left[
    \begin{array}{ccc}
      \tilde{\phi}_{\rm A} & \tilde{\phi}_{\rm B1} & \tilde{\phi}_{\rm B2} 
    \end{array}
  \right]^{\mathrm T}
=
U
\mathcal{P}(\bvec{k}_{||})
\left[
    \begin{array}{ccc}
      \tilde{q}_{\rm A} & \tilde{q}_{\rm B1} & \tilde{q}_{\rm B2} 
    \end{array}
  \right]^{\mathrm T}, 
\end{align}
where
\begin{align}\label{3}
\small
\mathcal{P}(\bvec{k}_{||})=
\eta
\left[
    \begin{array}{ccc}
      \eta^{-1} & 1+{\rm e}^{{\rm i}\bvec{k}_{||} \cdot \bvec{a}_{1}} & 1+{\rm e}^{-{\rm i}\bvec{k}_{||} \cdot \bvec{a}_{2}} \\
      1+{\rm e}^{-{\rm i}\bvec{k}_{||} \cdot \bvec{a}_{1}} & \eta^{-1} & 0 \\
      1+{\rm e}^{{\rm i}\bvec{k}_{||} \cdot \bvec{a}_{2}} & 0 & \eta^{-1}
    \end{array}
  \right], 
\end{align}
with $\eta = U'/U$. 
We use tildes to represent complex amplitudes, and a harmonic time dependence ${\rm e}^{-{\rm i}\omega t}$ with angular frequency $\omega$ is assumed. 

Next, we introduce the current $I_{ij}$ flowing from the $j$th disk to the $i$th disk and the self-inductance $L$ of the bar connecting the $i$th and $j$th disks. 
Then we have 
\begin{align}\label{4}
\cfrac{{\rm d}I_{ij}}{{\rm d}t}=-\cfrac{1}{L}(\phi_{i} - \phi_{j}). 
\end{align}
From Eq.~(\ref{4}) and the charge conservation law ${\rm d}q_{i}/{{\rm d}t}=\sum_{j}A_{ij}I_{ij}$ at the $i$th disk, we obtain
\begin{align}\label{5}
\cfrac{{\rm d}^2q_{i}}{{\rm d}t^2}=-\cfrac{1}{L}\sum_{j}A_{ij}(\phi_{i} - \phi_{j}). 
\end{align}
For the LBDRs, Eq.~(\ref{5}) is reduced to
\begin{align}\label{6}
\small
\cfrac{1}{L}
\mathcal{Q}(\bvec{k}_{||})
  \left[
    \begin{array}{ccc}
      \tilde{\phi}_{\rm A} & \tilde{\phi}_{\rm B1} & \tilde{\phi}_{\rm B2} 
    \end{array}
  \right]^{\mathrm T}
=
\omega^2
\left[
    \begin{array}{ccc}
      \tilde{q}_{\rm A} & \tilde{q}_{\rm B1} & \tilde{q}_{\rm B2} 
    \end{array}
  \right]^{\mathrm T} 
\end{align}
in the frequency domain. The matrix $\mathcal{Q}(\bvec{k}_{||})$ is defined as follows: 
\begin{align}\label{7}
\small
\mathcal{Q}(\bvec{k}_{||})=
-
\left[
    \begin{array}{ccc}
      -4 & 1+{\rm e}^{{\rm i}\bvec{k}_{||} \cdot \bvec{a}_{1}} & 1+{\rm e}^{-{\rm i}\bvec{k}_{||} \cdot \bvec{a}_{2}} \\
      1+{\rm e}^{-{\rm i}\bvec{k}_{||} \cdot \bvec{a}_{1}} & -2 & 0 \\
      1+{\rm e}^{{\rm i}\bvec{k}_{||} \cdot \bvec{a}_{2}} & 0 & -2
    \end{array}
  \right]. 
\end{align}
Equations (\ref{2}) and (\ref{6}) give the following eigenvalue equation: 
\begin{align}\label{8}
\mathcal{H}(\bvec{k}_{||})
\left[
    \begin{array}{ccc}
      \tilde{q}_{\rm A} & \tilde{q}_{\rm B1} & \tilde{q}_{\rm B2} 
    \end{array}
  \right]^{\mathrm T}
=
\left( \cfrac{\omega}{\omega_{0}} \right)^2
\left[
    \begin{array}{ccc}
      \tilde{q}_{\rm A} & \tilde{q}_{\rm B1} & \tilde{q}_{\rm B2} 
    \end{array}
  \right]^{\mathrm T}, 
\end{align}
where $\mathcal{H}(\bvec{k}_{||})=\mathcal{Q}(\bvec{k}_{||})\mathcal{P}(\bvec{k}_{||})$ and $\omega_{0}=\sqrt{U/L}$. 
Solving Eq.~(\ref{8}), we finally derive the dispersion relations 
 \begin{widetext}
   \vspace*{-5mm}
\begin{align}\label{9}
\cfrac{\omega}{\omega_{0}}=\sqrt{2},\, \sqrt{3-2\left[2+F(\bvec{k}_{||})\right]\eta \pm \sqrt{5-24\eta+32\eta^2+2\left( 1 -6\eta +8\eta^2 \right)F(\bvec{k}_{||})}}, 
\end{align}
 \vspace*{-5mm}
 \end{widetext}
where $F(\bvec{k}_{||})={{\rm cos}(\bvec{k}_{||} \cdot \bvec{a}_{1}) + {\rm cos}(\bvec{k}_{||} \cdot \bvec{a}_{2})}$. 
From these equations, we obtain the two-dimensional band structure shown in Fig.~\ref{fig1}(c) for $\eta = 0$. 
Note that $\eta=0$ corresponds to neglecting the nearest neighbor {\it capacitive} couplings between the disks, but the disks are  sufficiently coupled through currents flowing in the bars. 
There are three bands, and the middle band at $\omega/\omega_{0}=\sqrt{2}$ is flat. 
Although usual tight-binding models with the same on-site potential present the degeneracy of a flat band and one Dirac cone at the $\mathrm{M}$ point, only the lowest band touches the flat band at the ${\mathrm{M}}$ point in Fig.~\ref{fig1}(c). 
This is caused by on-site potential difference on the main lattice and sublattice for spoof plasmons on the Lieb lattice. 
From Eqs. (\ref{1}) and (\ref{5}) with $\eta=0$, we have
\vspace{-2mm}
\begin{align}\label{10}
\cfrac{{\mathrm{d}^{2}q_{i}}}{{\mathrm{d}}t^{2}}=\cfrac{U}{L}\left(\sum_{j}A_{ij}q_{j}\right)-\cfrac{U}{L}m_{i}q_{i}, 
\end{align}
where $m_{i}$ is the number of bars connected to the $i$th disk. 
In Eq.~(\ref{10}), the first term and the second term represent hopping and on-site potential, respectively. 
We can see that on-site potentials on the main lattice and sublattice are actually different in Eq.~(\ref{10}). 
In this study, we focus our attention on the flat band. 

The flat bands are formed by the localized modes with the charge distribution shown in Fig.~\ref{fig1}(d). 
In the flat-band mode, the charges oscillate in antiphase on the sublattice and no charges are induced on the main lattice owing to destructive interference. 
Hence, spoof plasmons in the flat band are two-dimensionally confined. 

\vspace{-5mm}
\section{Simulation}\label{sec3}
\vspace{-3mm}

To confirm the result of the previous theoretical calculation, we numerically simulate the electromagnetic response of the LBDRs. 
We use a commercial finite-element method solver ({Ansys HFSS}). 
The design parameters of the LBDRs in Fig.~\ref{fig1}(a) are as follows: the length of the unit cell is $a=420\;{\rm \mu m}$, the bar width is $w=27\;{\rm \mu m}$, and the disk radius is $r=76.5\;{\rm \mu m}$. 
In our simulation, the LBDRs are made of infinitely thin, perfect electric conductors. 
To observe the flat band both in the radiative region and in the
nonradiative region, we treat two cases: free-space transmission and
attenuated total reflection (ATR) configurations \cite{maier2007plasmonics}.
The transmission and ATR spectra, respectively, reflect the band structure of the LBDRs in the radiative and nonradiative regions. 

\begin{figure}[t]
\includegraphics[scale=0.24]{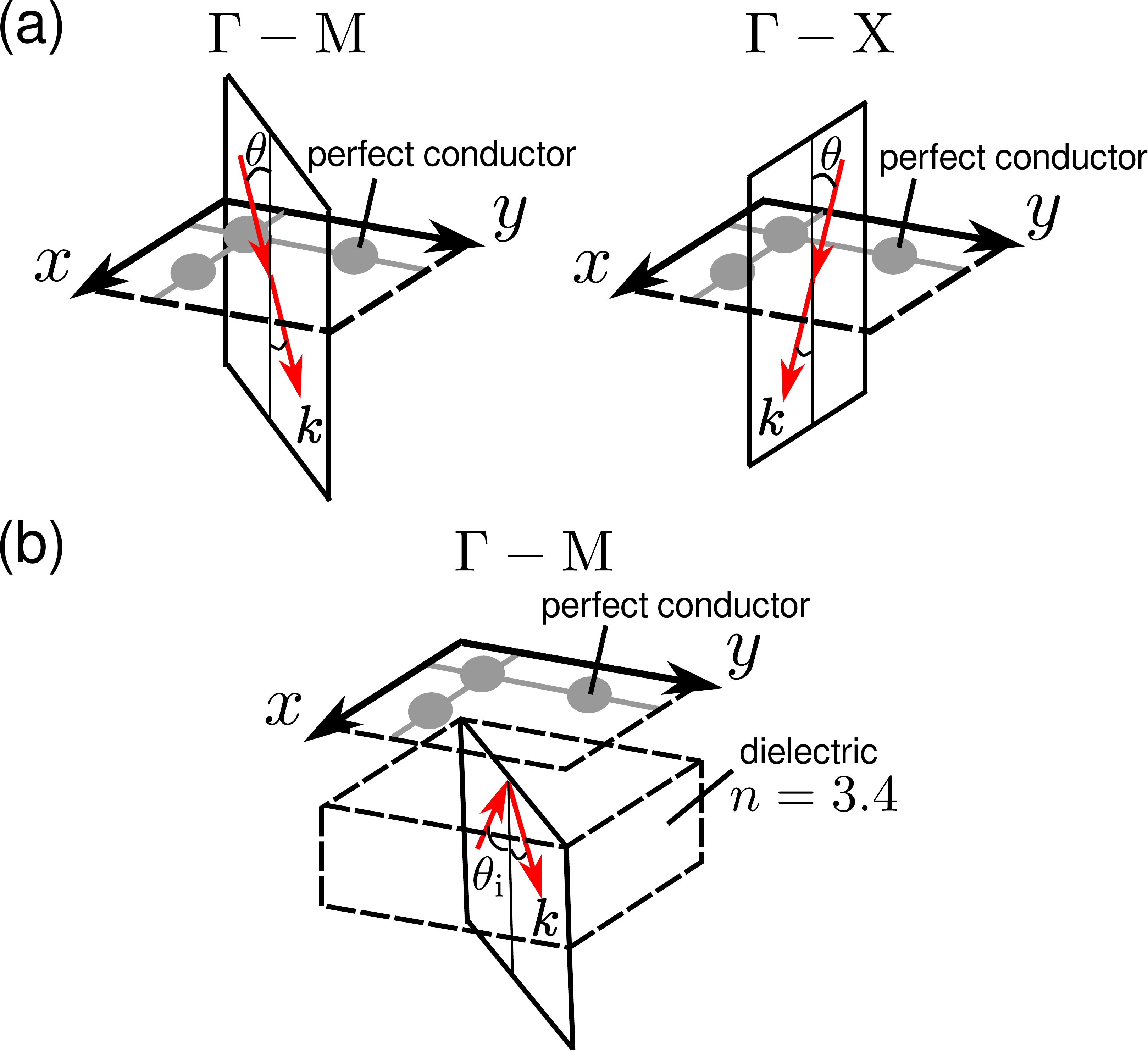}
\caption{\label{fig2} (Color online) Simulation setups. (a) Transmission simulation setup. A plane wave enters into the LBDRs at an incident angle $\theta$, which ranges from $\theta=0^{\circ}$ to $\theta=90^{\circ}$ with a step $\Delta \theta = 5^{\circ}$. The incident wave is in the TE mode for the $\Gamma$--$\mathrm{M}$ scan, and the TM mode for the $\Gamma$--$\mathrm{X}$ scan. (b)  ATR simulation setup. A plane wave enters from a dielectric, which has refractive index $n=3.4$, into the air. The incident angle $\theta_{\rm i}$ ranges from $\theta_{\rm i}=17.15^{\circ}$ to $\theta_{\rm i}=40^{\circ}$ ($\theta_{\rm i}=17.15^{\circ},\,17.5^{\circ}$, and $20^{\circ}$ to $40^{\circ}$ with a step $\Delta \theta_{\rm i}= 2^{\circ}$). The incident wave is in the TE mode for the $\Gamma$--$\mathrm{M}$ scan. }
\vspace{-3mm}
\end{figure}

\begin{figure}[h]
\includegraphics[scale=0.27]{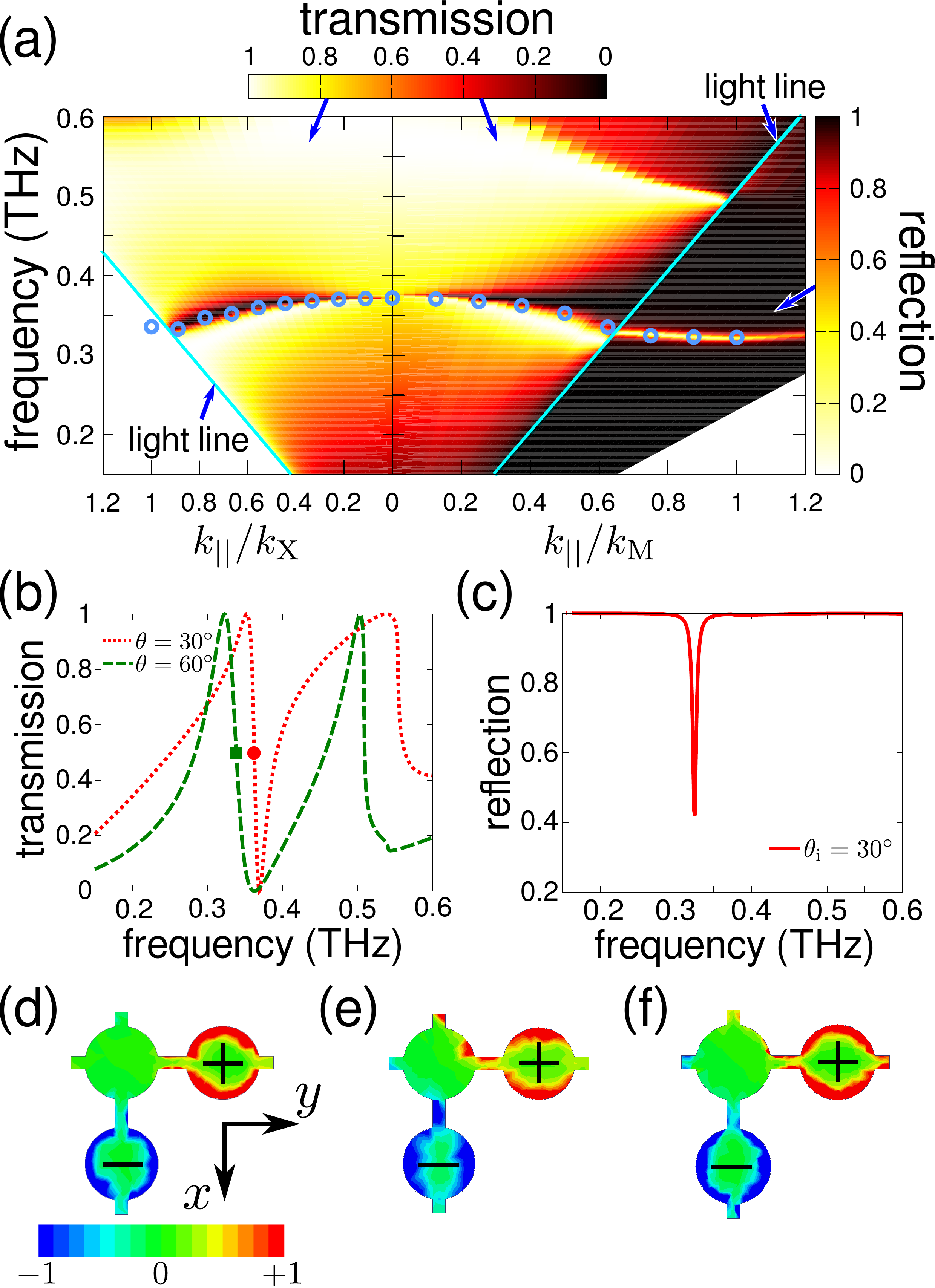}
\caption{\label{fig3} (Color online) Simulation results. (a) Power transmission and ATR power spectra mapped on the wave-vector--frequency plane. The open circles indicate eigenfrequencies calculated by the eigenmode solver of \textsc{HFSS}. 
The wave number $k_{||}$ is normalized by $k_{\mathrm M}=\sqrt{2}\pi/a$ and $k_{\mathrm{X}}=\pi/a$. 
The value $k_{\parallel}/k_{\mathrm M}=1$ corresponds to the ${\mathrm M}$ point, and the value $k_{\parallel}/k_{\mathrm X}=1$ corresponds to the ${\mathrm X}$ point. 
(b) Transmission spectra at  $\theta=30^{\circ}$ and $60^{\circ}$ in the $\Gamma$--$\mathrm{M}$ scan. 
The circle on the dotted line and the square on the dashed line correspond to the points where the transmission is 0.5. 
(c) ATR spectrum at $\theta_{\rm i}=30^{\circ}$ in the $\Gamma$--$\mathrm{M}$ scan. Surface electric charge distributions at 
(d) 0.362~THz for $\theta=30^{\circ}$ (the circle on the dotted line), (e) 0.338~THz for $\theta=60^{\circ}$ (the square on the dashed line), and (f) 0.322~THz for $\theta_{\rm i}=30^{\circ}$ (the dip in the ATR spectrum). 
}
\vspace{-3mm}
\end{figure}

Figure~\ref{fig2}(a) shows the setup of the transmission simulation. 
A plane wave enters into the LBDRs at an incident angle $\theta$, which is varied from $\theta=0^{\circ}$ to $90^{\circ}$. 
In this case, the magnitude of the wave vector $\bvec{k}_{||}$ on the sample plane is given by $k_{||}=(\omega/c)\sin \theta$, where $c$ is the speed of light in vacuum. 
In order to excite the flat band mode, the incident wave is in the transverse electric (TE) mode for the $\Gamma$--$\mathrm{M}$ scan, and the transverse magnetic (TM) mode for the $\Gamma$--$\mathrm{X}$ scan. 

Figure~\ref{fig2}(b) shows the setup of the ATR simulation. 
The distance between the LBDRs and the dielectric is set to $200\;{\rm \mu m}$ to reduce the proximity effect of the dielectric, as described in Sec~\ref{sec5}\@. 
A plane wave enters from the dielectric, which has refractive index $n$, into the air. 
If the incident angle $\theta_{\rm i}$ is larger than the critical angle $\theta_{0}={\rm arcsin}(1/n)$, the incident wave is totally reflected,  generating evanescent waves in the air. 
The refractive index of the dielectric is set to $n=3.4$, assuming the
use of silicon in the THz region \cite{grischkowsky1990far}. The incident angle $\theta_{\rm i}$ is varied from $\theta_{\rm i}=17.15^{\circ}$ to $40^{\circ}$, all of which are greater than $\theta_{0}=17.1^{\circ}$. 
In this case, the magnitude of the wave vector $\bvec{k}_{||}$ on the sample plane is given by $k_{||}=n(\omega/c)\sin \theta_{\rm i}$. 
The incident wave is in the TE mode for the $\Gamma$--$\mathrm{M}$ scan. 
The nonradiative region is negligibly small in the 1st Brillouin zone along the $\Gamma$--$\mathrm{X}$ line, so the ATR simulation is performed only in the $\Gamma$--$\mathrm{M}$ line.

By using periodic boundary conditions with phase shifts, the transmission,  reflection, and surface electric charge distribution in the unit cell are calculated for various incident angles. 
Figure~\ref{fig3}(a) displays the simulation results of the transmission and ATR spectra in the wave-vector--frequency plane. 
We can observe a narrow band insensitive to the wave vector in the nonradiative region. 
We also observe a broader band in the radiative region from $0.2$ to $0.6\;{\rm THz}$. 

To find where the flat band is located exactly, we use the eigenmode solver of \textsc{HFSS}, which provides the field distributions of resonant modes and their resonant frequencies for given structures and environments. 
In this simulation, we use periodic boundary conditions and the perfectly matched layer (PML). 
The PML regions are about half-wavelength apart from the LBDRs. 
In this configuration, the solver calculates the resonant mode propagating along the surface. 
The calculated eigenfrequencies are indicated by the open circles in Fig.~\ref{fig3}(a). 
The charge distribution of this eigenmode corresponds to the flat-band mode illustrated in Fig.~\ref{fig1}(d); therefore, we find that the narrow band in the nonradiative region is the flat band. 
On the other hand, the flat band in the radiative region is embedded in the high-transmission regions that are broadly distributed around $0.2$ to $0.6~{\rm THz}$ and is located on the boundary between the high- and low- transmission regions around $0.35~{\rm THz}$. 
It is revealed that the flat band is bending, and the degree of the bend is about $15\%$ of the bottom frequency of the flat band along the $\Gamma$--$\mathrm{M}$ line, and $11\%$ along the $\Gamma$--$\mathrm{X}$ line. The reason for the bend is discussed in Sec.~\ref{sec5}. 

Figure \ref{fig3}(b) shows the transmission spectra at incident angles $\theta=30^{\circ}$ and $\theta=60^{\circ}$ in the $\Gamma$--$\mathrm{M}$ scan, and Fig.~\ref{fig3}(c) shows the ATR spectrum at an incident angle $\theta_{\rm i}=30^{\circ}$ in the $\Gamma$--$\mathrm{M}$ scan. 
The ATR spectrum is simpler than the transmission spectra. 
For the transmission spectra, it is presumed that the flat-band mode
interferes with the broad mode, as is the case in Fano interference \cite{fano1961effects}.

Figures \ref{fig3}(d) and \ref{fig3}(e) show the surface electric charge distributions in the transmission configuration at the resonant frequencies: 0.362\;THz for $\theta=30^{\circ}$ and 0.338\;THz for $\theta=60^{\circ}$ in the $\Gamma$--$\mathrm{M}$ scan. 
Figure \ref{fig3}(f) shows the surface electric charge distributions of the ATR configuration at the resonant frequency 0.322\;THz for $\theta_{\rm i}=30^{\circ}$ in the $\Gamma$--$\mathrm{M}$ scan. 
It is clear that the surface charge oscillates in antiphase on the two disks of the sublattice in the unit cell. 
This resonant mode coincides with the theoretical charge distribution of the flat-band mode described in Sec.~\ref{sec2}. 
Therefore, we can conclude that the flat band is observed. 

We also calculate the group velocity $v_{\rm g}$ and the quality factor of the flat-band mode in the nonradiative region from the eigenfrequencies obtained by the eigenmode solver along the $\Gamma$--$\mathrm{M}$ line. 
The group velocity $v_{\rm g} \sim -c/60$ is obtained from the slope of the straight line fitted to the eigenfrequencies in the nonradiative region. 
In our simulation, we obtain the quality factor $Q_{\rm out} = 1.7 \times 10^4$ in the nonradiative region, which is a factor of $\sim$350 higher than the quality factor $Q_{\rm in}=48$ in the radiative region. 

  \begin{figure}[]
\includegraphics[scale=0.30]{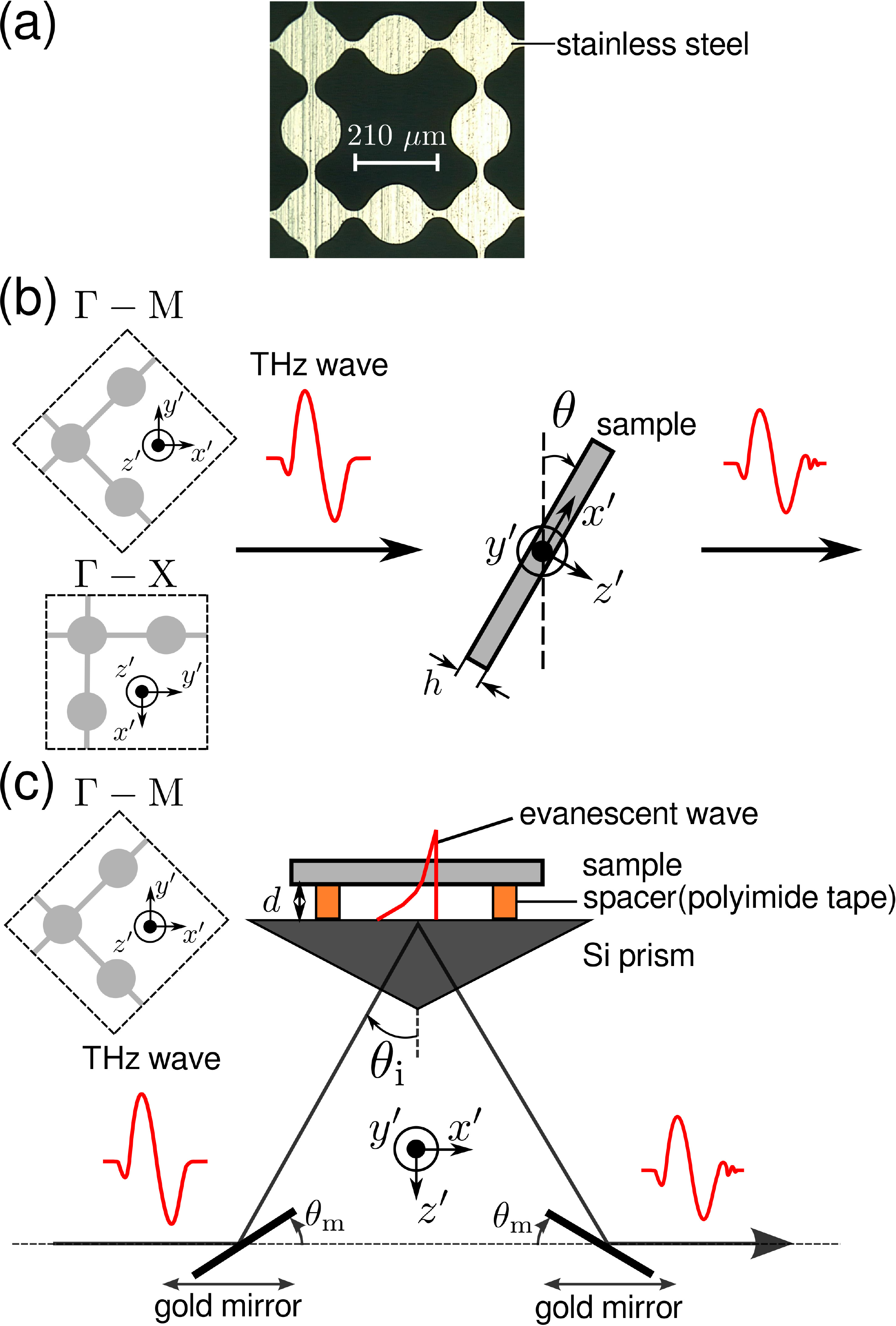}
\caption{\label{fig4} (Color online) (a) Photomicrograph of the LBDRs
 fabricated on a stainless-steel (SUS304) sheet with thickness
 $h=30\;{\rm \mu m}$. (b) Schematic of the transmission measurement with
 THz-TDS. (c) Schematic of the total reflection measurement with THz
 TD-ATR spectroscopy.  The distance between the LBDRs and the prism is
 $d \sim 64\;{\rm \mu m}$. }
 \vspace{-3mm}
\end{figure}

\begin{figure}[t]
\includegraphics[scale=0.3]{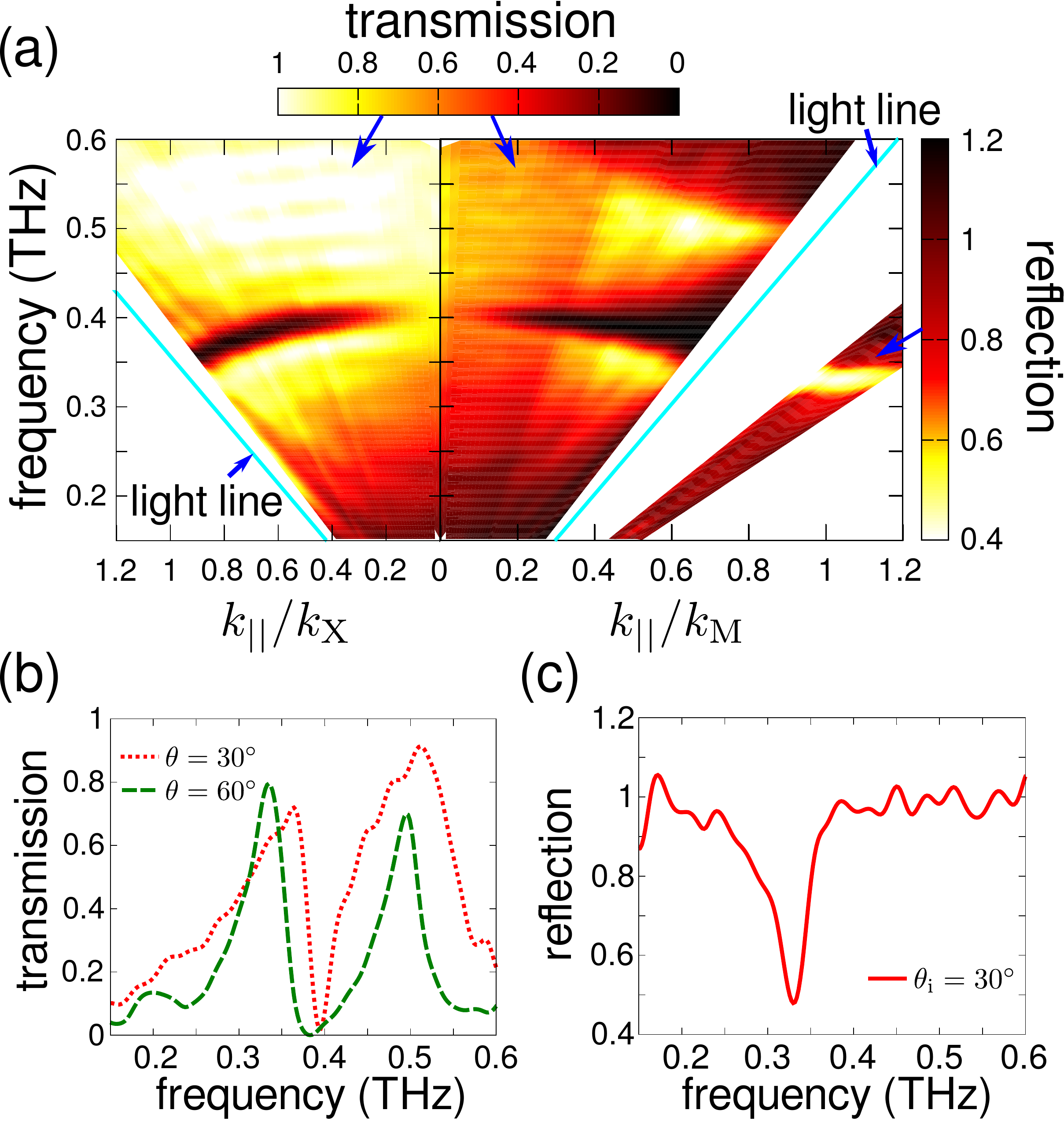}
\caption{\label{fig5} (Color online) Experimental results. (a)
 Transmission and ATR spectra mapped on the wave-vector--frequency
 plane. The value $k_{\parallel}/k_{\mathrm M}=1$ on the horizontal axis
 corresponds to the ${\mathrm M}$ point, where $k_{\mathrm
 M}=\sqrt{2}\pi/a$. The value $k_{\parallel}/k_{\mathrm X}=1$ on the
 horizontal axis corresponds to the ${\mathrm X}$ point, where
 $k_{\mathrm X}=\pi/a$. (b) Transmission spectra for the incident angles
 $\theta=30^{\circ}$ and $60^{\circ}$ in the $\Gamma$--$\mathrm{M}$
 scan. (c)  ATR spectrum for the incident angle $\theta_{\rm
 i}=30^{\circ}$ in the $\Gamma$--$\mathrm{M}$ scan. }
 \vspace{-2mm}
\end{figure}

\vspace{-5mm}
\section{Experiments}\label{sec4}
\vspace{-2mm}
  \subsection{Experimental setup}
\vspace{-2mm}

  Figure~\ref{fig4}(a) shows the LBDRs patterned on a stainless-steel (SUS304) sheet with thickness $h=30\;{\rm \mu m}$ by etching. 
The period of the lattice is $420\;{\rm \mu m}$ and the whole area is $1.1 \times 1.1\;{\rm cm}^2$. 
To observe the dispersion relation in the radiative and nonradiative
regions experimentally, we perform a transmission measurement using THz
time-domain spectroscopy (THz-TDS), as shown in Fig.~\ref{fig4}(b)
\cite{nakata2012observation}, and a total reflection measurement using
THz time-domain attenuated total reflection (TD-ATR) spectroscopy, as
shown in Fig.~\ref{fig4}(c)
\cite{hirori2004attenuated}.
We use a THz emitter and detector ({EKSPLA, Ltd.}) with dipole antennas. 
The antennas are fixed on low-temperature-grown GaAs photoconductors illuminated by a femtosecond fiber laser (F-100, {IMRA America, Inc.}) with a wavelength of $\lambda=810\;{\rm nm}$ and pulse duration of  $\tau=120\;{\rm fs}$. 
A silicon lens is used to produce a collimated THz beam with diameter of $7.4\;{\rm mm}$. 
The electric field of the THz wave, $E(t),$ is measured in the time domain. 
In the TD-ATR measurement, a silicon prism \cite{grischkowsky1990far} with refractive index $n=3.4$ is used. 
The distance between the LBDRs and the prism is $d \sim 64\;{\rm \mu m}$ (the thickness of the polyimide spacing tape). 
We can obtain the transmission spectrum $T(\omega)$ and the ATR spectrum $R(\omega)$ from $T(\omega)=|\tilde{E}(\omega)/\tilde{E}_{\rm ref}(\omega)|^2$ and $R(\omega)=|\tilde{E}(\omega)/\tilde{E}_{\rm ref}(\omega)|^2$, where $\tilde{E}(\omega)$ and $\tilde{E}_{\rm ref}(\omega)$ are the Fourier-transformed electric fields with and without the sample, respectively.

To observe the band structure experimentally, the incident angle is changed in each measurement as follows: 
(a) In the transmission measurement, the sample is rotated by $\theta$ around the $y'$ axis from normal incidence, as shown in Fig.~\ref{fig4}(b). 
The incident angle $\theta$ is changed from $\theta=0^{\circ}$ to $65^{\circ}$ in steps of $\Delta \theta = 2.5^{\circ}$, and the corresponding wave number is given by $k_{||}=(\omega/c)\sin \theta$. 
The incident wave is in the TE mode for the $\Gamma$--$\mathrm{M}$ scan, and the TM mode for the $\Gamma$--$\mathrm{X}$ scan. 
(b) In the total reflection measurement, the gold mirrors are rotated from $\theta_{\rm m}=28^{\circ}$ to $38^{\circ}$ by $\Delta \theta_{\rm m}=2^{\circ}$, corresponding to changing the incident angle from $\theta_{\rm i} = 25.3^{\circ}$ to $\theta_{\rm i} = 31.2^{\circ}$, as shown in Fig.~\ref{fig4}(c). 
The incident wave is in the TE mode. 
The angle, or in-plane wave vector, is limited by the finite size of the mirror mounts and the finite area of the LBDRs. 
The corresponding wave number is given by $k_{||}=n(\omega/c)\sin \theta_{\rm i}$.

\vspace{-3mm}
\subsection{Results}
\vspace{-2mm}

Figure~\ref{fig5} displays the results of the experiments. 
The transmission and ATR spectra mapped on the wave-vector--frequency plane are shown in Fig.~\ref{fig5}(a). 
In the nonradiative region, a single ATR dip that is insensitive to the wave number is observed. 
In the radiative region, there exists a broader band from $0.2$ to $0.6\;{\rm THz}$. 
These results are qualitatively and almost quantitatively in good agreement with the simulation shown in Fig.~\ref{fig3}(a). 
Figure~\ref{fig5}(b) shows the transmission spectra at the incident angles $\theta=30^{\circ}$ and $\theta=60^{\circ}$ in the $\Gamma$--$\mathrm{M}$ scan. 
Figure~\ref{fig5}(c) shows the ATR spectrum at the incident angle $\theta_{\rm i}=30^{\circ}$ in the $\Gamma$--$\mathrm{M}$ scan. 
These results also agree well with the results of the simulation.

\begin{figure}[b]
\includegraphics[scale=0.3]{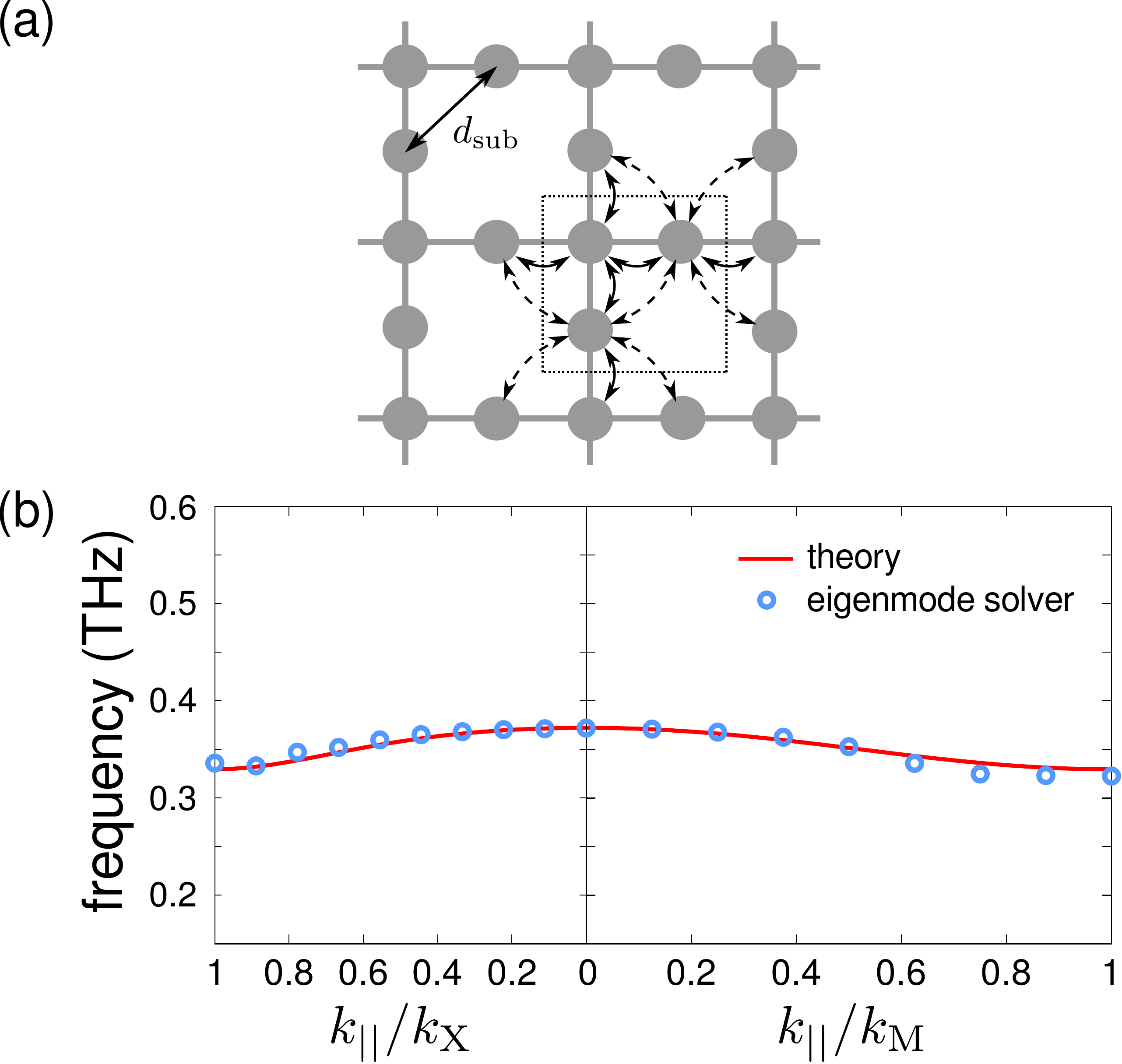}
\caption{\label{fig6} (Color online) (a) Capacitive couplings between the nearest-neighbor disks (solid arrows) and the second-nearest-neighbor disks (dashed arrows). (b) Theoretical curve (solid line) fitted to the eigenfrequencies calculated by the eigenmode solver. The parameters are $\omega_{0}/(2\pi)=0.233\;{\rm THz}$, $\eta=0,$ and $\eta'=-0.0690$. }
\vspace{-3mm}
\end{figure}

Compared with the simulation result obtained for ideal conditions, the resonant frequency of the flat-band mode is a little higher, and the line width of the spectrum broadens. 
The possible reasons for these disagreements are as follows: 
First, although the LBDRs are assumed to be perfect conductors in the simulation, the fabricated LBDRs used in the experiment have the finite conductivity of the stainless steel. 
It has been confirmed that the line width of the spectrum broadens also in the simulation for the LBDRs with finite conductivity (see Sec.\;\ref{sec5}). 
Second, owing to the lack of precision of the etching, the width of the bar is widened. 
This causes a decrease of the effective inductance of the bar, and the resonant frequency shifts higher. 
Finally, the distance between the LBDRs and the prism, $d \sim 64\;{\rm \mu m,}$ in the experiment differs from the simulation parameter $d=200\;{\rm \mu m}$. 
The dependence of the ATR spectra on the distance between the LBDRs and the prism is discussed in Sec.~\ref{sec5}.

\vspace{-2mm}
\section{Discussion}\label{sec5}
\vspace{-2mm}
  \subsection{Origin of the bend of the flat band}
\vspace{-2mm}

The flat-band frequency is slightly dependent on the wave vector both in the simulation and in the experiment. 
As a major cause we consider the second-nearest-neighbor capacitive couplings, as shown in Fig.~\ref{fig6}(a), in addition to the nearest-neighbor couplings. 
In this model, Eq.~(\ref{1}) is replaced with  
\begin{align}\label{11}
\phi_{i} = Uq_{i}+U'\sum_{j}A_{ij}q_{j} + U''\sum_{k}B_{ik}q_{k}. 
\end{align}
We define the distance between the centers of the two disks of the sublattice in a unit cell as $d_{\rm sub}$. 
$B_{ik}=1$ if the $k$th disk exists within a radius $d_{\rm sub}$ from
the center of the $i$th disk; otherwise $B_{ik}=0$.

By using Eq.~(\ref{11}) instead of Eq.~(\ref{1}), $\mathcal{P}(\bvec{k}_{||})$ in Eq.~(\ref{2}) is replaced with 
 \begin{widetext}
  \vspace{-3mm}
\begin{align}\label{12}
\mathcal{P'}(\bvec{k}_{||})=
\mathcal{P}(\bvec{k}_{||})+
\eta'
\left[
    \begin{array}{ccc}
      0 & 0 &\hspace{-5pt} 0 \\
      0 & 0 &\hspace{-5pt} 1+{\rm e}^{-{\rm i}\bvec{k}_{||} \cdot \bvec{a}_{1}}+{\rm e}^{-{\rm i}\bvec{k}_{||} \cdot \bvec{a}_{2}}+{\rm e}^{-{\rm i}\bvec{k}_{||} \cdot (\bvec{a}_{1}+\bvec{a}_{2})} \hspace{-5pt} \\
      0 & 1+{\rm e}^{{\rm i}\bvec{k}_{||} \cdot \bvec{a}_{1}}+{\rm e}^{{\rm i}\bvec{k}_{||} \cdot \bvec{a}_{2}}+{\rm e}^{{\rm i}\bvec{k}_{||} \cdot (\bvec{a}_{1}+\bvec{a}_{2})} &\hspace{-5pt} 0 
    \end{array}
  \right], 
\end{align}
 \end{widetext}
where $\eta'=U''/U$. 
The dispersion relation is obtained by solving the eigenvalue problem  $\mathcal{H'}(\bvec{k}_{||})=\mathcal{Q}(\bvec{k}_{||})\mathcal{P'}(\bvec{k}_{||})$.

The solid line in Fig.~\ref{fig6}(b) is the theoretical curve fitted to the eigenfrequencies (shown as open circles) that are obtained by the eigenmode solver. 
By solving the eigenvalue problem of $\mathcal{H'}(\bvec{k}_{||})$, we obtain the frequencies of the flat-band mode at the $\Gamma$ point as $\omega_{{\rm f},\,\Gamma}/(2\pi)=\sqrt{2-8\eta'}\omega_{0}/(2\pi)$, the ${\mathrm M}$ point as $\omega_{{\rm f},\,{\mathrm M}}/(2\pi)=\sqrt{2}\omega_{0}/(2\pi)$, and the $\mathrm{X}$ point as $\omega_{{\rm f},\,{\mathrm X}}/(2\pi)=\sqrt{2}\omega_{0}/(2\pi)$. 
By averaging of the eigenfrequencies $0.323$ and $0.336\;{\rm THz}$ at the $\mathrm{M}$ and $\mathrm{X}$ points, the parameter $\omega_{0}/(2\pi)=0.233\:{\rm THz}$ is determined. 
From the eigenfrequency $0.372\;{\rm THz}$ at the $\Gamma$ point, the parameter $\eta'=-0.0690$ is determined. 
We note that the variation of the curve is affected  only by the parameter $\eta'$, not by $\eta$. 
This is because the second-nearest-neighbor couplings break the condition on the formation of the localized modes. 
For simplification, we assume the value of $\eta$ as $0$. If $\eta' > 0$, the band takes its minimal value at $\bvec{k}_{||}=0$; in contrast, if $\eta' < 0$, it takes its maximal value. 
In the static limit ($\omega \to 0$), positive charges on a disk create positive electric potential on the other disks; hence the parameter $\eta'$ is expected to be positive. 
However, the actual value of $\eta'$ is negative. 
This can be explained by the retardation effect \cite{liu2012optical, tatartschuk2012mapping}.
The phase difference between the second-nearest disks is given by $(\sqrt{2}\omega_{0}/c) \times a/\sqrt{2} \sim 0.65 \times \pi$ at $\sqrt{2}\omega_{0}/(2\pi)=0.330\;{\rm THz}$. 
The negative sign of $\eta'$ is caused by this shift being greater than $\pi/2$. 
Note that the second-nearest-neighbor coupling is weakened by reducing the disk radius and the bar width without changing the length of the unit cell, and we could further suppress the degree of the bend by improving the precision of the fabrication. 

\begin{figure}[b]
\includegraphics[scale=0.26]{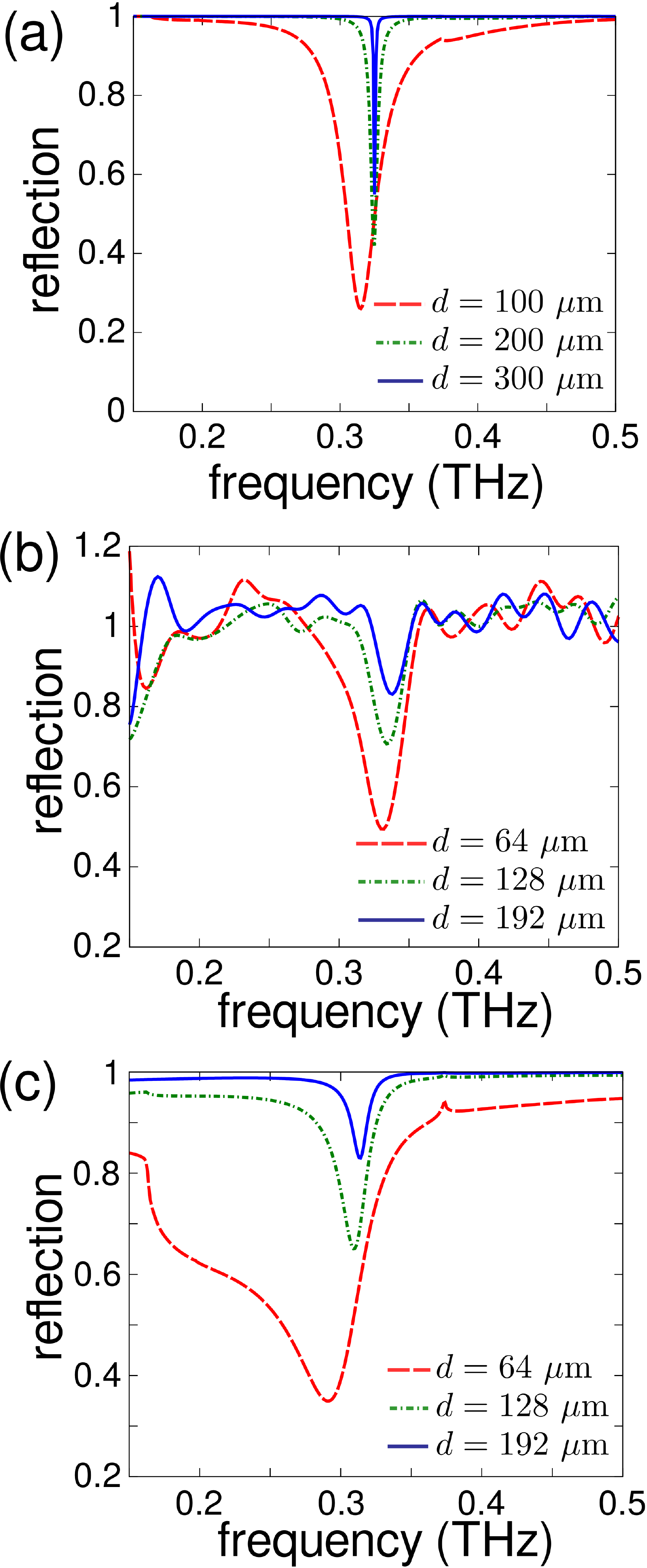}
\caption{\label{fig7} (Color online) Dependence of the ATR spectra on the distance between the LBDRs and the prism. (a) Simulation results of the ATR spectra of the perfectly conducting LBDRs. (b) Experimental results of the ATR spectra. (c) Simulation results of the ATR spectra in the case of the LBDRs with the DC conductivity of SUS304. }
\end{figure}

\subsection{Dependence of ATR spectra on the distance between the LBDRs and the prism}
The flat band in the nonradiative region cannot couple with free-space modes; theoretically, this mode has no energy radiation into free space and is perfectly confined at the surface. 
However, in our situation, the nonradiative flat-band mode is excited by evanescent waves. 
The evanescent waves are generated by the total reflection of the propagating waves; that is, the flat-band mode couples with the propagating waves. 
This coupling strength depends on the distance $d$ between the LBDRs and the prism. 
It is expected that the width of the ATR dip becomes narrower by increasing the distance $d$ because the coupling strength is weakened. 

To verify the coupling effect, we perform total reflection measurements with several distances in the simulation and the experiment. 
The simulation is performed for $d=100$, 200, and $300\;{\rm \mu m}$ under the same conditions of the ATR configuration in Sec.~\ref{sec3}. 
In the experiment, the distance $d$ is changed by increasing the number of  polyimide spacer layers from one to three. 
Figures~\ref{fig7}(a) and \ref{fig7}(b) show the results of the simulation and the experiment, respectively. 
These results demonstrate  that the ATR dip becomes narrower and shallower as $d$ increases. 
Furthermore, we see that, as the distance between the LBDRs and the prism is increased, the resonant frequency shifts to higher frequency and converges to the resonant frequency in free space. 
Figure\;\ref{fig7}(c) shows the ATR spectra obtained by the simulation
in the case where the finite-conductivity boundary condition with the DC
conductivity $\sigma = 1.4 \times 10^6 {\rm \;S/m}$ of SUS304 is applied
to the infinitely thin LBDRs region \cite{davis1998metals}.
From this result, we can confirm that the spectra are broadened owing to the finite conductivity in the ATR experiment. 

\vspace{-4mm}
\section{Conclusion}\label{sec6}
\vspace{-2mm}
In this paper, we demonstrated the three-dimensionally confined flat-band mode for spoof surface plasmons. 
We analytically calculated the dispersion relation of spoof plasmons on the LBDRs based on the electrical circuit model. 
We obtained three bands, the middle of which is a flat band. 
To confirm this theoretical result, we numerically simulated the electromagnetic response of the LBDRs using the finite-element method. 
The simulation results revealed that the flat band in the nonradiative
region shows narrower line width than that in the radiative region. 
Besides, we found that the flat-band bend can be explained by the second-nearest-neighbor capacitive coupling. 
Finally, we performed  transmission  and  total reflection measurements using the THz-TDS and the TD-ATR methods. 
The experimental results showed qualitatively good agreement with the simulation results. 

In the total reflection measurement, as the distance between the sample and the coupling prism is increased, the coupling strength between the spoof surface plasmons and the evanescent wave is weakened, and the ATR dip becomes narrower. 
Because spatially localized modes can be formed, these modes can be
applied to sensitive plasmon sensors
\cite{yoshida2007terahertz} with high spatial resolution. 
Furthermore, combined with the low group velocity, the flat-band mode
with a high quality factor can be used to enhance the nonlinear response
of materials \cite{PhysRevB.92.125124}.

%
%
\begin{acknowledgments}
 We gratefully thank Dr.~Atsushi Yao and Professor Takashi Hikihara for allowing us to use their equipment and Professor Fumiaki Miyamaru for fruitful discussions. 
The present research is supported by JSPS KAKENHI Grants No. 22109004, No. 25790065, and No. 25287101.
\end{acknowledgments}
%
%

%
%
\bibliographystyle{apsrev4-1}

\begin{thebibliography}{25}%
\makeatletter
\providecommand \@ifxundefined [1]{%
 \@ifx{#1\undefined}
}%
\providecommand \@ifnum [1]{%
 \ifnum #1\expandafter \@firstoftwo
 \else \expandafter \@secondoftwo
 \fi
}%
\providecommand \@ifx [1]{%
 \ifx #1\expandafter \@firstoftwo
 \else \expandafter \@secondoftwo
 \fi
}%
\providecommand \natexlab [1]{#1}%
\providecommand \enquote  [1]{``#1''}%
\providecommand \bibnamefont  [1]{#1}%
\providecommand \bibfnamefont [1]{#1}%
\providecommand \citenamefont [1]{#1}%
\providecommand \href@noop [0]{\@secondoftwo}%
\providecommand \href [0]{\begingroup \@sanitize@url \@href}%
\providecommand \@href[1]{\@@startlink{#1}\@@href}%
\providecommand \@@href[1]{\endgroup#1\@@endlink}%
\providecommand \@sanitize@url [0]{\catcode `\\12\catcode `\$12\catcode
  `\&12\catcode `\#12\catcode `\^12\catcode `\_12\catcode `\%12\relax}%
\providecommand \@@startlink[1]{}%
\providecommand \@@endlink[0]{}%
\providecommand \url  [0]{\begingroup\@sanitize@url \@url }%
\providecommand \@url [1]{\endgroup\@href {#1}{\urlprefix }}%
\providecommand \urlprefix  [0]{URL }%
\providecommand \Eprint [0]{\href }%
\providecommand \doibase [0]{http://dx.doi.org/}%
\providecommand \selectlanguage [0]{\@gobble}%
\providecommand \bibinfo  [0]{\@secondoftwo}%
\providecommand \bibfield  [0]{\@secondoftwo}%
\providecommand \translation [1]{[#1]}%
\providecommand \BibitemOpen [0]{}%
\providecommand \bibitemStop [0]{}%
\providecommand \bibitemNoStop [0]{.\EOS\space}%
\providecommand \EOS [0]{\spacefactor3000\relax}%
\providecommand \BibitemShut  [1]{\csname bibitem#1\endcsname}%
\let\auto@bib@innerbib\@empty
\bibitem [{\citenamefont {Maier}(2007)}]{maier2007plasmonics}%
  \BibitemOpen
  \bibfield  {author} {\bibinfo {author} {\bibfnamefont {S.~A.}\ \bibnamefont
  {Maier}},\ }\href@noop {} {\emph {\bibinfo {title} {Plasmonics: fundamentals
  and applications}}}\ (\bibinfo  {publisher} {Springer Science \& Business
  Media, New York},\ \bibinfo {year} {2007})\BibitemShut {NoStop}%
\bibitem [{\citenamefont {Maier}\ \emph {et~al.}(2006)\citenamefont {Maier},
  \citenamefont {Andrews}, \citenamefont {Martin-Moreno},\ and\ \citenamefont
  {Garcia-Vidal}}]{maier2006terahertz}%
  \BibitemOpen
  \bibfield  {author} {\bibinfo {author} {\bibfnamefont {S.~A.}\ \bibnamefont
  {Maier}}, \bibinfo {author} {\bibfnamefont {S.~R.}\ \bibnamefont {Andrews}},
  \bibinfo {author} {\bibfnamefont {L.}~\bibnamefont {Martin-Moreno}}, \ and\
  \bibinfo {author} {\bibfnamefont {F.~J.}\ \bibnamefont {Garcia-Vidal}},\
  }\href@noop {} {\bibfield  {journal} {\bibinfo  {journal} {Phys. Rev. Lett.}\
  }\textbf {\bibinfo {volume} {97}},\ \bibinfo {pages} {176805} (\bibinfo
  {year} {2006})}\ \BibitemShut {NoStop}%
\bibitem [{\citenamefont {Williams}\ \emph {et~al.}(2008)\citenamefont
  {Williams}, \citenamefont {Andrews}, \citenamefont {Maier}, \citenamefont
  {Fern{\'a}ndez-Dom{\'\i}nguez}, \citenamefont {Mart{\'\i}n-Moreno},\ and\
  \citenamefont {Garcia-Vidal}}]{williams2008highly}%
  \BibitemOpen
  \bibfield  {author} {\bibinfo {author} {\bibfnamefont {C.~R.}\ \bibnamefont
  {Williams}}, \bibinfo {author} {\bibfnamefont {S.~R.}\ \bibnamefont
  {Andrews}}, \bibinfo {author} {\bibfnamefont {S.}~\bibnamefont {Maier}},
  \bibinfo {author} {\bibfnamefont {A.}~\bibnamefont
  {Fern{\'a}ndez-Dom{\'\i}nguez}}, \bibinfo {author} {\bibfnamefont
  {L.}~\bibnamefont {Mart{\'\i}n-Moreno}}, \ and\ \bibinfo {author}
  {\bibfnamefont {F.}~\bibnamefont {Garcia-Vidal}},\ }\href@noop {} {\bibfield
  {journal} {\bibinfo  {journal} {Nature Photon.}\ }\textbf {\bibinfo {volume}
  {2}},\ \bibinfo {pages} {175} (\bibinfo {year} {2008})}\ \BibitemShut
  {NoStop}%
\bibitem [{\citenamefont {Pendry}\ \emph {et~al.}(2004)\citenamefont {Pendry},
  \citenamefont {Martin-Moreno},\ and\ \citenamefont
  {Garcia-Vidal}}]{pendry2004mimicking}%
  \BibitemOpen
  \bibfield  {author} {\bibinfo {author} {\bibfnamefont {J.~B.}\ \bibnamefont
  {Pendry}}, \bibinfo {author} {\bibfnamefont {L.}~\bibnamefont
  {Martin-Moreno}}, \ and\ \bibinfo {author} {\bibfnamefont {F.~J.}\
  \bibnamefont {Garcia-Vidal}},\ }\href@noop {} {\bibfield  {journal} {\bibinfo
   {journal} {Science}\ }\textbf {\bibinfo {volume} {305}},\ \bibinfo {pages}
  {847} (\bibinfo {year} {2004})}\ \BibitemShut {NoStop}%
\bibitem [{\citenamefont {Fern{$\acute{\rm a}$}ndez-Dom{\'\i}nguez}\ \emph
  {et~al.}(2008)\citenamefont {Fern{$\acute{\rm a}$}ndez-Dom{\'\i}nguez},
  \citenamefont {Mart{\'\i}n-Moreno}, \citenamefont {Garc{\'\i}a-Vidal},
  \citenamefont {Andrews},\ and\ \citenamefont {Maier}}]{martin2008spoof}%
  \BibitemOpen
  \bibfield  {author} {\bibinfo {author} {\bibfnamefont {A.~I.}\ \bibnamefont
  {Fern{$\acute{\rm a}$}ndez-Dom{\'\i}nguez}}, \bibinfo {author} {\bibfnamefont
  {L.}~\bibnamefont {Mart{\'\i}n-Moreno}}, \bibinfo {author} {\bibfnamefont
  {F.~J.}\ \bibnamefont {Garc{\'\i}a-Vidal}}, \bibinfo {author} {\bibfnamefont
  {S.~R.}\ \bibnamefont {Andrews}}, \ and\ \bibinfo {author} {\bibfnamefont
  {S.~A.}\ \bibnamefont {Maier}},\ }\href@noop {} {\bibfield  {journal}
  {\bibinfo  {journal} {IEEE J. Sel. Topics in Quantum Electron.}\ }\textbf
  {\bibinfo {volume} {14}},\ \bibinfo {pages} {1515} (\bibinfo {year} {2008})}\
  \BibitemShut {NoStop}%
\bibitem [{\citenamefont {Gan}\ \emph {et~al.}(2008)\citenamefont {Gan},
  \citenamefont {Fu}, \citenamefont {Ding},\ and\ \citenamefont
  {Bartoli}}]{gan2008ultrawide}%
  \BibitemOpen
  \bibfield  {author} {\bibinfo {author} {\bibfnamefont {Q.}~\bibnamefont
  {Gan}}, \bibinfo {author} {\bibfnamefont {Z.}~\bibnamefont {Fu}}, \bibinfo
  {author} {\bibfnamefont {Y.~J.}\ \bibnamefont {Ding}}, \ and\ \bibinfo
  {author} {\bibfnamefont {F.~J.}\ \bibnamefont {Bartoli}},\ }\href@noop {}
  {\bibfield  {journal} {\bibinfo  {journal} {Phys. Rev. Lett.}\ }\textbf
  {\bibinfo {volume} {100}},\ \bibinfo {pages} {256803} (\bibinfo {year}
  {2008})}\ \BibitemShut {NoStop}%
\bibitem [{\citenamefont {Yu}\ \emph {et~al.}(2010)\citenamefont {Yu},
  \citenamefont {Wang}, \citenamefont {Kats}, \citenamefont {Fan},
  \citenamefont {Khanna}, \citenamefont {Li}, \citenamefont {Davies},
  \citenamefont {Linfield},\ and\ \citenamefont {Capasso}}]{yu2010designer}%
  \BibitemOpen
  \bibfield  {author} {\bibinfo {author} {\bibfnamefont {N.}~\bibnamefont
  {Yu}}, \bibinfo {author} {\bibfnamefont {Q.~J.}\ \bibnamefont {Wang}},
  \bibinfo {author} {\bibfnamefont {M.~A.}\ \bibnamefont {Kats}}, \bibinfo
  {author} {\bibfnamefont {J.~A.}\ \bibnamefont {Fan}}, \bibinfo {author}
  {\bibfnamefont {S.~P.}\ \bibnamefont {Khanna}}, \bibinfo {author}
  {\bibfnamefont {L.}~\bibnamefont {Li}}, \bibinfo {author} {\bibfnamefont
  {A.~G.}\ \bibnamefont {Davies}}, \bibinfo {author} {\bibfnamefont {E.~H.}\
  \bibnamefont {Linfield}}, \ and\ \bibinfo {author} {\bibfnamefont
  {F.}~\bibnamefont {Capasso}},\ }\href@noop {} {\bibfield  {journal} {\bibinfo
   {journal} {Nature Mater.}\ }\textbf {\bibinfo {volume} {9}},\ \bibinfo
  {pages} {730} (\bibinfo {year} {2010})}\ \BibitemShut {NoStop}%
\bibitem [{\citenamefont {Lieb}(1989)}]{PhysRevLett.62.1201}%
  \BibitemOpen
  \bibfield  {author} {\bibinfo {author} {\bibfnamefont {E.~H.}\ \bibnamefont
  {Lieb}},\ }\href {\doibase 10.1103/PhysRevLett.62.1201} {\bibfield  {journal}
  {\bibinfo  {journal} {Phys. Rev. Lett.}\ }\textbf {\bibinfo {volume} {62}},\
  \bibinfo {pages} {1201} (\bibinfo {year} {1989})}\ \BibitemShut {NoStop}%
\bibitem [{\citenamefont {Tasaki}(1992)}]{tasaki1992ferromagnetism}%
  \BibitemOpen
  \bibfield  {author} {\bibinfo {author} {\bibfnamefont {H.}~\bibnamefont
  {Tasaki}},\ }\href@noop {} {\bibfield  {journal} {\bibinfo  {journal} {Phys.
  Rev. Lett.}\ }\textbf {\bibinfo {volume} {69}},\ \bibinfo {pages} {1608}
  (\bibinfo {year} {1992})}\ \BibitemShut {NoStop}%
\bibitem [{\citenamefont {Mielke}(1991)}]{mielke1991ferromagnetism}%
  \BibitemOpen
  \bibfield  {author} {\bibinfo {author} {\bibfnamefont {A.}~\bibnamefont
  {Mielke}},\ }\href@noop {} {\bibfield  {journal} {\bibinfo  {journal} {J.
  Phys. A Math Gen.}\ }\textbf {\bibinfo {volume} {24}},\ \bibinfo {pages}
  {3311} (\bibinfo {year} {1991})}\ \BibitemShut {NoStop}%
\bibitem [{\citenamefont {Takeda}\ \emph {et~al.}(2004)\citenamefont {Takeda},
  \citenamefont {Takashima},\ and\ \citenamefont {Yoshino}}]{takeda2004flat}%
  \BibitemOpen
  \bibfield  {author} {\bibinfo {author} {\bibfnamefont {H.}~\bibnamefont
  {Takeda}}, \bibinfo {author} {\bibfnamefont {T.}~\bibnamefont {Takashima}}, \
  and\ \bibinfo {author} {\bibfnamefont {K.}~\bibnamefont {Yoshino}},\
  }\href@noop {} {\bibfield  {journal} {\bibinfo  {journal} {J. Phys. Condens.
  Matter}\ }\textbf {\bibinfo {volume} {16}},\ \bibinfo {pages} {6317}
  (\bibinfo {year} {2004})}\ \BibitemShut {NoStop}%
\bibitem [{\citenamefont {Vicencio}\ \emph {et~al.}(2015)\citenamefont
  {Vicencio}, \citenamefont {Cantillano}, \citenamefont {Morales-Inostroza},
  \citenamefont {Real}, \citenamefont {Mej{\'\i}a-Cort{\'e}s}, \citenamefont
  {Weimann}, \citenamefont {Szameit},\ and\ \citenamefont
  {Molina}}]{vicencio2015observation}%
  \BibitemOpen
  \bibfield  {author} {\bibinfo {author} {\bibfnamefont {R.~A.}\ \bibnamefont
  {Vicencio}}, \bibinfo {author} {\bibfnamefont {C.}~\bibnamefont
  {Cantillano}}, \bibinfo {author} {\bibfnamefont {L.}~\bibnamefont
  {Morales-Inostroza}}, \bibinfo {author} {\bibfnamefont {B.}~\bibnamefont
  {Real}}, \bibinfo {author} {\bibfnamefont {C.}~\bibnamefont
  {Mej{\'\i}a-Cort{\'e}s}}, \bibinfo {author} {\bibfnamefont {S.}~\bibnamefont
  {Weimann}}, \bibinfo {author} {\bibfnamefont {A.}~\bibnamefont {Szameit}}, \
  and\ \bibinfo {author} {\bibfnamefont {M.~I.}\ \bibnamefont {Molina}},\
  }\href@noop {} {\bibfield  {journal} {\bibinfo  {journal} {Phys. Rev. Lett.}\
  }\textbf {\bibinfo {volume} {114}},\ \bibinfo {pages} {245503} (\bibinfo
  {year} {2015})}\ \BibitemShut {NoStop}%
\bibitem [{\citenamefont {Mukherjee}\ \emph {et~al.}(2015)\citenamefont
  {Mukherjee}, \citenamefont {Spracklen}, \citenamefont {Choudhury},
  \citenamefont {Goldman}, \citenamefont {\"Ohberg}, \citenamefont
  {Andersson},\ and\ \citenamefont {Thomson}}]{PhysRevLett.114.245504}%
  \BibitemOpen
  \bibfield  {author} {\bibinfo {author} {\bibfnamefont {S.}~\bibnamefont
  {Mukherjee}}, \bibinfo {author} {\bibfnamefont {A.}~\bibnamefont
  {Spracklen}}, \bibinfo {author} {\bibfnamefont {D.}~\bibnamefont
  {Choudhury}}, \bibinfo {author} {\bibfnamefont {N.}~\bibnamefont {Goldman}},
  \bibinfo {author} {\bibfnamefont {P.}~\bibnamefont {\"Ohberg}}, \bibinfo
  {author} {\bibfnamefont {E.}~\bibnamefont {Andersson}}, \ and\ \bibinfo
  {author} {\bibfnamefont {R.~R.}\ \bibnamefont {Thomson}},\ }\href {\doibase
  10.1103/PhysRevLett.114.245504} {\bibfield  {journal} {\bibinfo  {journal}
  {Phys. Rev. Lett.}\ }\textbf {\bibinfo {volume} {114}},\ \bibinfo {pages}
  {245504} (\bibinfo {year} {2015})}\ \BibitemShut {NoStop}%
\bibitem [{\citenamefont {Endo}\ \emph {et~al.}(2010)\citenamefont {Endo},
  \citenamefont {Oka},\ and\ \citenamefont {Aoki}}]{endo2010tight}%
  \BibitemOpen
  \bibfield  {author} {\bibinfo {author} {\bibfnamefont {S.}~\bibnamefont
  {Endo}}, \bibinfo {author} {\bibfnamefont {T.}~\bibnamefont {Oka}}, \ and\
  \bibinfo {author} {\bibfnamefont {H.}~\bibnamefont {Aoki}},\ }\href@noop {}
  {\bibfield  {journal} {\bibinfo  {journal} {Phys. Rev. B}\ }\textbf {\bibinfo
  {volume} {81}},\ \bibinfo {pages} {113104} (\bibinfo {year} {2010})}\
  \BibitemShut {NoStop}%
\bibitem [{\citenamefont {Feigenbaum}\ and\ \citenamefont
  {Atwater}(2010)}]{feigenbaum2010resonant}%
  \BibitemOpen
  \bibfield  {author} {\bibinfo {author} {\bibfnamefont {E.}~\bibnamefont
  {Feigenbaum}}\ and\ \bibinfo {author} {\bibfnamefont {H.~A.}\ \bibnamefont
  {Atwater}},\ }\href@noop {} {\bibfield  {journal} {\bibinfo  {journal} {Phys.
  Rev. Lett.}\ }\textbf {\bibinfo {volume} {104}},\ \bibinfo {pages} {147402}
  (\bibinfo {year} {2010})}\ \BibitemShut {NoStop}%
\bibitem [{\citenamefont {Nakata}\ \emph
  {et~al.}(2012{\natexlab{a}})\citenamefont {Nakata}, \citenamefont {Okada},
  \citenamefont {Nakanishi},\ and\ \citenamefont {Kitano}}]{nakata2012circuit}%
  \BibitemOpen
  \bibfield  {author} {\bibinfo {author} {\bibfnamefont {Y.}~\bibnamefont
  {Nakata}}, \bibinfo {author} {\bibfnamefont {T.}~\bibnamefont {Okada}},
  \bibinfo {author} {\bibfnamefont {T.}~\bibnamefont {Nakanishi}}, \ and\
  \bibinfo {author} {\bibfnamefont {M.}~\bibnamefont {Kitano}},\ }\href@noop {}
  {\bibfield  {journal} {\bibinfo  {journal} {Phys. Status Solidi (b)}\
  }\textbf {\bibinfo {volume} {249}},\ \bibinfo {pages} {2293} (\bibinfo {year}
  {2012}{\natexlab{a}})}\ \BibitemShut {NoStop}%
\bibitem [{\citenamefont {Nakata}\ \emph
  {et~al.}(2012{\natexlab{b}})\citenamefont {Nakata}, \citenamefont {Okada},
  \citenamefont {Nakanishi},\ and\ \citenamefont
  {Kitano}}]{nakata2012observation}%
  \BibitemOpen
  \bibfield  {author} {\bibinfo {author} {\bibfnamefont {Y.}~\bibnamefont
  {Nakata}}, \bibinfo {author} {\bibfnamefont {T.}~\bibnamefont {Okada}},
  \bibinfo {author} {\bibfnamefont {T.}~\bibnamefont {Nakanishi}}, \ and\
  \bibinfo {author} {\bibfnamefont {M.}~\bibnamefont {Kitano}},\ }\href@noop {}
  {\bibfield  {journal} {\bibinfo  {journal} {Phys. Rev. B}\ }\textbf {\bibinfo
  {volume} {85}},\ \bibinfo {pages} {205128} (\bibinfo {year}
  {2012}{\natexlab{b}})}\ \BibitemShut {NoStop}%
\bibitem [{\citenamefont {Grischkowsky}\ \emph {et~al.}(1990)\citenamefont
  {Grischkowsky}, \citenamefont {Keiding}, \citenamefont {Van~Exter},\ and\
  \citenamefont {Fattinger}}]{grischkowsky1990far}%
  \BibitemOpen
  \bibfield  {author} {\bibinfo {author} {\bibfnamefont {D.}~\bibnamefont
  {Grischkowsky}}, \bibinfo {author} {\bibfnamefont {S.}~\bibnamefont
  {Keiding}}, \bibinfo {author} {\bibfnamefont {M.}~\bibnamefont {Van~Exter}},
  \ and\ \bibinfo {author} {\bibfnamefont {C.}~\bibnamefont {Fattinger}},\
  }\href@noop {} {\bibfield  {journal} {\bibinfo  {journal} {J. Opt. Soc. Am.
  B}\ }\textbf {\bibinfo {volume} {7}},\ \bibinfo {pages} {2006} (\bibinfo
  {year} {1990})}\ \BibitemShut {NoStop}%
\bibitem [{\citenamefont {Fano}(1961)}]{fano1961effects}%
  \BibitemOpen
  \bibfield  {author} {\bibinfo {author} {\bibfnamefont {U.}~\bibnamefont
  {Fano}},\ }\href@noop {} {\bibfield  {journal} {\bibinfo  {journal} {Phys.
  Rev.}\ }\textbf {\bibinfo {volume} {124}},\ \bibinfo {pages} {1866} (\bibinfo
  {year} {1961})}\ \BibitemShut {NoStop}%
\bibitem [{\citenamefont {Hirori}\ \emph {et~al.}(2004)\citenamefont {Hirori},
  \citenamefont {Yamashita}, \citenamefont {Nagai},\ and\ \citenamefont
  {Tanaka}}]{hirori2004attenuated}%
  \BibitemOpen
  \bibfield  {author} {\bibinfo {author} {\bibfnamefont {H.}~\bibnamefont
  {Hirori}}, \bibinfo {author} {\bibfnamefont {K.}~\bibnamefont {Yamashita}},
  \bibinfo {author} {\bibfnamefont {M.}~\bibnamefont {Nagai}}, \ and\ \bibinfo
  {author} {\bibfnamefont {K.}~\bibnamefont {Tanaka}},\ }\href@noop {}
  {\bibfield  {journal} {\bibinfo  {journal} {Jpn. J. Appl. Phys.}\ }\textbf
  {\bibinfo {volume} {43}},\ \bibinfo {pages} {L1287} (\bibinfo {year}
  {2004})}\ \BibitemShut {NoStop}%
\bibitem [{\citenamefont {Liu}\ \emph {et~al.}(2012)\citenamefont {Liu},
  \citenamefont {Powell}, \citenamefont {Shadrivov},\ and\ \citenamefont
  {Kivshar}}]{liu2012optical}%
  \BibitemOpen
  \bibfield  {author} {\bibinfo {author} {\bibfnamefont {M.}~\bibnamefont
  {Liu}}, \bibinfo {author} {\bibfnamefont {D.~A.}\ \bibnamefont {Powell}},
  \bibinfo {author} {\bibfnamefont {I.~V.}\ \bibnamefont {Shadrivov}}, \ and\
  \bibinfo {author} {\bibfnamefont {Y.~S.}\ \bibnamefont {Kivshar}},\
  }\href@noop {} {\bibfield  {journal} {\bibinfo  {journal} {Appl. Phys.
  Lett.}\ }\textbf {\bibinfo {volume} {100}},\ \bibinfo {pages} {111114}
  (\bibinfo {year} {2012})}\ \BibitemShut {NoStop}%
\bibitem [{\citenamefont {Tatartschuk}\ \emph {et~al.}(2012)\citenamefont
  {Tatartschuk}, \citenamefont {Gneiding}, \citenamefont {Hesmer},
  \citenamefont {Radkovskaya},\ and\ \citenamefont
  {Shamonina}}]{tatartschuk2012mapping}%
  \BibitemOpen
  \bibfield  {author} {\bibinfo {author} {\bibfnamefont {E.}~\bibnamefont
  {Tatartschuk}}, \bibinfo {author} {\bibfnamefont {N.}~\bibnamefont
  {Gneiding}}, \bibinfo {author} {\bibfnamefont {F.}~\bibnamefont {Hesmer}},
  \bibinfo {author} {\bibfnamefont {A.}~\bibnamefont {Radkovskaya}}, \ and\
  \bibinfo {author} {\bibfnamefont {E.}~\bibnamefont {Shamonina}},\ }\href@noop
  {} {\bibfield  {journal} {\bibinfo  {journal} {J. Appl. Phys.}\ }\textbf
  {\bibinfo {volume} {111}},\ \bibinfo {pages} {094904} (\bibinfo {year}
  {2012})}\ \BibitemShut {NoStop}%
\bibitem [{\citenamefont {Davis}(1998)}]{davis1998metals}%
  \BibitemOpen
  \bibfield  {author} {\bibinfo {author} {\bibfnamefont {J.~R.}\ \bibnamefont
  {Davis}},\ }\href@noop {} {\emph {\bibinfo {title} {Metals handbook}}}\
  (\bibinfo  {publisher} {ASM international, Materials Park, OH},\ \bibinfo
  {year} {1998})\BibitemShut {NoStop}%
\bibitem [{\citenamefont {Yoshida}\ \emph {et~al.}(2007)\citenamefont
  {Yoshida}, \citenamefont {Ogawa}, \citenamefont {Kawai}, \citenamefont
  {Hayashi}, \citenamefont {Hayashi}, \citenamefont {Otani}, \citenamefont
  {Kato}, \citenamefont {Miyamaru},\ and\ \citenamefont
  {Kawase}}]{yoshida2007terahertz}%
  \BibitemOpen
  \bibfield  {author} {\bibinfo {author} {\bibfnamefont {H.}~\bibnamefont
  {Yoshida}}, \bibinfo {author} {\bibfnamefont {Y.}~\bibnamefont {Ogawa}},
  \bibinfo {author} {\bibfnamefont {Y.}~\bibnamefont {Kawai}}, \bibinfo
  {author} {\bibfnamefont {S.}~\bibnamefont {Hayashi}}, \bibinfo {author}
  {\bibfnamefont {A.}~\bibnamefont {Hayashi}}, \bibinfo {author} {\bibfnamefont
  {C.}~\bibnamefont {Otani}}, \bibinfo {author} {\bibfnamefont
  {E.}~\bibnamefont {Kato}}, \bibinfo {author} {\bibfnamefont {F.}~\bibnamefont
  {Miyamaru}}, \ and\ \bibinfo {author} {\bibfnamefont {K.}~\bibnamefont
  {Kawase}},\ }\href@noop {} {\bibfield  {journal} {\bibinfo  {journal}
  {Applied physics letters}\ }\textbf {\bibinfo {volume} {91}},\ \bibinfo
  {pages} {253901} (\bibinfo {year} {2007})}\ \BibitemShut {NoStop}%
\bibitem [{\citenamefont {Tamayama}\ \emph {et~al.}(2015)\citenamefont
  {Tamayama}, \citenamefont {Hamada},\ and\ \citenamefont
  {Yasui}}]{PhysRevB.92.125124}%
  \BibitemOpen
  \bibfield  {author} {\bibinfo {author} {\bibfnamefont {Y.}~\bibnamefont
  {Tamayama}}, \bibinfo {author} {\bibfnamefont {K.}~\bibnamefont {Hamada}}, \
  and\ \bibinfo {author} {\bibfnamefont {K.}~\bibnamefont {Yasui}},\ }\href
  {\doibase 10.1103/PhysRevB.92.125124} {\bibfield  {journal} {\bibinfo
  {journal} {Phys. Rev. B}\ }\textbf {\bibinfo {volume} {92}},\ \bibinfo
  {pages} {125124} (\bibinfo {year} {2015})}\ \BibitemShut {NoStop}%
\end{thebibliography}

%

\end{document}